# Cold Supply Chain Planning including Smart Contracts: An Intelligent Blockchain-based approach

Soroush Goodarzi
*Department of Industrial Engineering*
*Sharif University of Technology*
Tehran, Iran
soroush.goodarzi@ie.sharif.edu

Vahid Kayvanfar
*Department of Industrial Engineering*
*Amirkabir University of Technology*
Tehran, Iran

Alireza Haji
*Department of Industrial Engineering*
*Sharif University of Technology*
Tehran, Iran
ahaji@sharif.edu

Alireza Shirzad
*Department of Electrical Engineering*
*Sharif University of Technology*
Tehran, Iran
alireza.shirzad7@student.sharif.edu

*Abstract*— Vaccinating the global population against Covid-19 is one of the biggest supply chain management challenges humanity has ever faced. Rapid supply of Covid-19 vaccines is essential for successful global immunization, but its effectiveness depends on a transparent supply chain that can be monitored. In this research, we have proposed an approach based on blockchain technology, which is used to ensure seamless distribution of the Covid-19 vaccine with transparency, data integrity, and full traceability of the supply chain to reduce risk, ensure safety, and immutability. A vaccine supply chain needs to update the status of the vaccine at every stage, and any problem in the supply and distribution path can lead to irreparable damage. Currently, the research conducted on the use of blockchain in supply chains is still in the early stages. In this paper, the use of blockchain technology to monitor the vaccine supply and distribution system will be investigated. A model close to reality of today's vaccine supply chains in developing countries is considered and then a new intelligent system for vaccine monitoring in the vaccine supply chain is designed based on the considered model. Also, smart contracts based on a blockchain network is designed to check consumer vaccination records as well as vaccine circulation from beginning to end. The implementation and design of the vaccine supply chain is done using smart contracts on the Ethereum blockchain network. Additionally, the system has been tested on both local networks, the HardHat suite and Rinkbey's test network. The system has also been developed to work seamlessly when it is using an integrated IoT chip that can automatically update a batch's location, temperature, and other physical conditions periodically.

## I. Introduction

Immunization with a vaccine is an essential process in any modern society to prevent infectious and viral diseases. Therefore, it directly affects public health and national security. Today, the spread of the Covid-19 pandemic has increased the significance of this process.

So far immunization with vaccines is known to be one of the most effective and economical ways to maintain public health and saves two to three millions lives each year. In addition, new vaccines such as pneumococcal conjugate vaccine (PCV) and rotavirus vaccine can save even one million more lives annually. Yet, general immunization programs are facing challenges in terms of introducing new vaccines, providing a stable budget, as well as vaccine supply and distribution.

Some new vaccines are being developed and distributed all around the world (and some other vaccines are still in the production process). The difficulties and challenges that these new vaccines are faced with include: A) the risk of decreasing the effectiveness of the prescribed vaccines (for instance, due to lack of temperature control and proper equipment), B) The presence of fake vaccines in the system C)unavailability of immunization resources ( due to inadequate storage capacity, disruption in service delivery, vaccine storages, etc.) D) inefficient use of limited financial and human resources (for instance, as a consequence of wasting vaccines). Cold supply chain management can be utilized to solve some of these problems. Therefore, improving chain supply systems can effectively increase immunization coverage and decrease the fatality rate of preventable diseases.

Blockchain technology is a distributed database of public or private records or ledgers of digital events (transactions) that have been shared and executed among the participants. The basis of blockchain technology is distributed ledger technology. Blockchain technology has four key characteristics that distinguish it from other available information systems. These characteristics include lack of localization (decentralization), security, controllability and auditability, smart implementation (smart contract)

In Blockchain, a participant creates a new transaction to add it to the Blockchain. The new transaction is distributed throughout the network to be verified and audited. When the majority of blockchain nodes verify the transaction (according to predefined rules), the new transaction is added to the chain as a new block. A transaction record is distributed and saved among several nodes to ensure security. At the same time, smart contract, as one of the most important features of blockchain, provides valid transactions without requiring the involvement of third parties. Decentralization is another significant feature of Blockchain technology that detects and prevents fraud in information and consequently increases the validity of the data.

Blockchains are capable of changing the design, organization, operations and general management of supply chains. The fact that blockchain technology ensures security, traceability and authenticity of information and enables the automation of relationships by smart contract in an unreliable space all lead to a need for significant reconsideration in supply chains and their management.

Problems such as expired and fake vaccines are still prevalent in supply chains. Therefore, an effective management system is needed for monitoring vaccine supply chains. Subsequent to the emergence of the internet, blockchain technology is considered a starting point for developing next-generation technologies and is capable of altering current information management systems. Meanwhile, the development of machine learning technology is also providing other methods for analyzing data in information management systems. Together, blockchain and machine learning technology have a significant role in overcoming the challenges that vaccine supply chains face.

Today Blockchain is considered an outstanding technology which is changing previous technologies by reshaping traditional business models and creating new opportunities all over supply chain. In addition, blockchain enables traceability in supply chain management and establishes closer relationships and trust not only among organizations and suppliers but also across the entire supply chain. On one side, on the account of decentralization, smart contract can remarkably increase efficiency. On the other side, blockchain can work with other major technologies (such as big data analytics, internet of things, cyber-physical system, etc.) to reveal its evolutionary effect in all specialized fields.

## II. Previous work

Emergency supply chains are one of the most studied topics. For instance, Dwivedi and Shareef (2018) [1] in a study investigating emergency supply chain management in developing countries, mentioned vaccine supply. In another study, , et al. (2018) [2] aimed to understand complex humanitarian operations. also, Shareef et al. (2019)[3] discussed the problems and shortcomings of emergency supply chain and suggested developing an efficient emergency supply chain for managing emergencies.

More and more researchers are realizing that the advantages of using blockchain technology are not limited to cryptocurrency, and it can be beneficial in information management as well. For instance, Li, Kang, et al. (2018)[4] by implementing blockchain technology, investigated security and privacy issues caused by unreliable and intransparent energy markets and proposed a secure energy trading system. The result of their study indicated that energy blockchain is also efficient in the industrial internet of things. Herbaut & Negru (2017)[5] analyzed a method of implementing blockchain-powered smart contracts and network service chaining for supporting a user-centric content delivery ecosystem. Siokorski, Haughton & Kraft (2017)[6] studied the application of blockchain technology in the fourth industrial revolution. Thakur, Doja, Dwivedi, Ahmad & Khadanga (2020)[7] studied the application of blockchain technology in land records. There are more studies on the application of blockchain and machine learning technology in the vaccine supply chain that will be covered in the following pages.

There are studies on blockchains related to the supply chain. Tian (2016)[8] combined Radio Frequency Identification (RFID) and blockchain technology in order to develop a traceability system for the agricultural food supply chain. Lu and Xu (2017)[9] proposed the use of blockchain technology for tracing products through all stages of the supply chain. Queiroz and Wamba (2019)[10] studied the use of blockchain in India and the US and implement blockchain technology in the supply chain. Queiroz, Telles & Bonilla (2020)[11] did a similar research as well. For the purpose of inclusion of Indian people, who live in the rural areas, in the global supply chain, Schuetz & Venkatesh (2020)[12] proposed the adoption of blockchain technology in order to decrease poverty in the Villages of India. Tönnissen & Teuteberg (2020)[13] analyzed the effects of blockchain technology in supply chain management based on several previous studies. Behnke & Janssen (2020)[14] studied traceability in the food supply chain using blockchain technology and investigated its boundary conditions as well. Wong et al. (2020)[15] studied the effects of blockchain technology in supply chain management of small and medium enterprises in addition to the main factors that contribute to blockchain technology adoption in small and medium enterprises in Malesia. These studies strongly indicate that blockchain technology can be easily used to monitor supply chains and overcome challenges of vaccine safety as a result. In a recent study, Dolgui et al. (2020)[16] introduced a Blockchain-oriented dynamic model as well as a dynamic approach for solving the problem of smart contract design in a supply chain. Zhu & Kouhizadeh (2019)[17] analyzed the integration of supply chain management based on blockchain technology and its effects on related decisions. Kamble et al. (2020)[18] used blockchain technology for traceability in the agriculture supply chain. Their study encouraged other researchers to adopt blockchain technology for establishing relationships in the supply chain. Saberi, Kouhizadeh, et al. (2019)[19] analyzed programs that are based on blockchain and smart contracts in order to decrease fraud in supply chain management. All mentioned studies suggest that blockchain technology provides a promising future for implementing traceability and management functions in information management systems in the product supply chain. therefore, blockchain technology may solve vaccine safety challenges to some extent.

There are significant research gaps in this field. Blockchain technology has drawn the attention of researchers and scholars of various fields in recent years and created many opportunities and challenges. Although blockchain technology has been developed remarkably in recent years, as indicated in the review of literature, it has a noticeable gap when it comes to experimental studies. a considerable number of papers were published on blockchains in 2018 and 2019. But are not enough studies on blockchains related to healthcare information management. There is a noticeable research gap with regard to supply chain management and implementation, especially vaccine and medicine supply chain. Although blockchain technology is beneficial to supply chain management, only a few studies have been done on the adoption of blockchain for monitoring safe vaccination in supply chains. Although, the literature on blockchain-based supply chains still has a long way to go, it is a reliable basis for future studies. As mentioned earlier, most of these studies were done in developed countries, thus, there is a research gap with regard to such studies in developing countries. Moreover, the integration and interaction between blockchain technology and other modern technologies such as artificial intelligence and machine learning is another important topic that was not been fully addressed so far.

## III. OUR CONTRIBUTION

Although research on blockchain is still in its early steps, it is a reliable basis for information management studies. Interaction between blockchain and other advanced technologies can also potentially provide solutions for overcoming problems in operation and supply chain management. The available studies suggest that the United States, China, England, Germany, and South Korea are the top five countries that has referred the most to the related research, it can be concluded that such studies has been done in the most advanced and industrial countries. Therefore, it is of high importance for both academic researchers and industrial sections to implement such projects in the developing countries. Therefore, there is plenty of room for further research on blockchain technology in supply chain management.

The main purposes of this study are as follows:

- Addressing problems in vaccinations and creating a blockchain-based management system, in other words, "vaccine blockchain" to trace and manage information in vaccine supply chain

- Designing a system that enables vaccine traceability and smart contract operations based on a model close to reality that can prevent vaccine expiration and related frauds

- Also it is hoped that the findings of this study help other supply chains(especially drug supply chains) to overcome challenges in their management and safety monitoring

- The Main purpose of this research is to introduce a model for maintaining the records of the vaccine supply chain using the development of a system based on blockchain technology, while explaining this technology. Additionally, blockchain platforms and their dependencies are clearly explained.

The review of literature suggests that the current studies on application of blockchain in supply chains are still in early stages. This study aims to investigate application of blockchain technology in monitoring vaccine supply and distribution systems. Subsequently, a smart system is designed in order to monitor vaccine supply chain. Moreover, a blockchain-based smart contract is designed to trace vaccination records of the patients, as well as, the distribution of vaccines in vaccine production institutes, government, and health care centers. In addition, a new smart system for is designed to identify expired vaccines, as well as a system to automatically warn healthcare organization and supply chain monitoring organizations of expired vaccines. Finally, by using experimental results the ability of blockchain technology for ensuring validity of vaccine records are analyzed.

## IV. SYSTEM MODEL

Our system model is a simplified view of the real-world vaccine supply chain scenarios. This system is composed of 4 parties that do not trust each other, namely Authority, Transporter, Distributer and Vaccinator. There follows a short description of each entity:

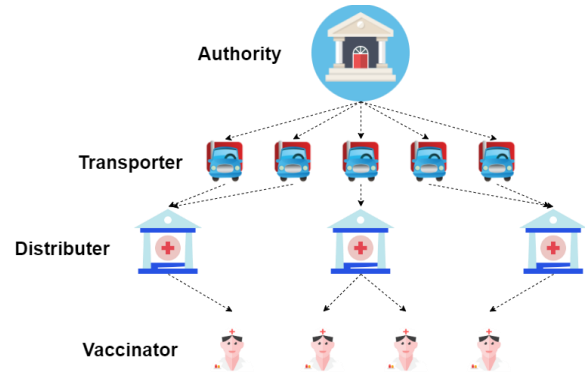

*Figure 1: System model*

- Authority is the first entity following the manufacturer of the vaccine. It is modeled after the highest authoritative department of a country responsible for public health, e.g. health ministry.

- Transporter is an entity responsible for transporting the vaccines throughout the country. This entity is composed of a lot of vehicles, storage rooms and refrigerators that move the vaccines between parties.

- Distributer is the local health departments placed in cities or villages to give people the basic health services.

- Vaccinator is a party that injects the vaccine to the patient.

Note that each of the above parties can be realized manually or in an automatic way. For example the transporter party can be the truck drivers or the devices embedded in the truck refrigerators.

## V. IMPLEMENTATION

This system has been implemented [1] and tested using solidity language (v 0.8.0) which is a high-level language that operates on an Ethereum virtual machine (EVM). All implementations are done using hardhat smart contract development toolkit. We chose Ethereum as the blockchain platform because it is one of the most studied and widely accepted Blockchains that supports smart contract development.

The implementation consists of two smart contracts that separate the system logic, namely AccessControl.sol and VaccineSupply.sol. The reason behind this separation was the ease of development and management of the contract.

### A. Access Control

In the Access Control smart contract, all the entities and owners and their access to the vaccines are controlled. This is done by the usage of modifiers and special data structures. First of all, each address is assigned an owner type that specifies the role of the address, i.e. whether It is an Authority or Distributer or, …. These roles can only be set by the

---

[1] https://github.com/alireza-shirzad/Vaccination_Blockchain

Authority. The Authority addresses are set at the deployment time by calling the smart contract constructor.

Another functionality of the Access Control smart contract is keeping ownership data of each vaccine, in which for each vaccine the record consists of the current owner's address and the next owner's address. The current owner's address points to the party that currently has possession of the vaccine. The next owner address is usually the zero address (null address) except at the time of handover which will be discussed in the Vaccine supply contract description.

about to be delivered to another party. The receiving party after the proper examination of the vaccine can either *accept* or *reject* this request. If accepted, The ownership will be fully transferred to the receiving party, If not, the ownership is returned to the initiator of the request.

In any part of the chain, if any party detects that the vaccine is compromised or expired, it can call the *expire* function to set the validity flag of the vaccine to *false*. This detection can be done manually or by digital sensors placed in the storages or refrigerators. We suggest that this process be done by the IOT sensors connected to the blockchain network.

At the end of the chain an *inject* call can be triggered by the *Vaccinator*, after the vaccine has been injected for a patient. The patient then needs to confirm this injection by the *patience_receive_vaccine* call that terminates the lifecycle of a vaccine.

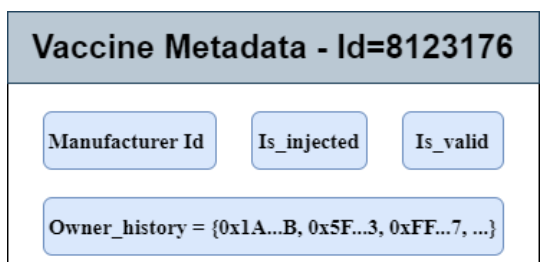

*Figure 2: Access Control main mappings*

## VI. TEST AND DEPLOYMENT

The smart contract has been tested using the Hardhat framework using numerous unit tests. It was deployed both on the Hardhat network which is an emulator of the blockchain and on the Rinkeby test network using Infura APIs.

### B. Vaccine Supply

This contract has the core functionalities of the supply chain. Each vaccine is given an id to be tracked during the supply process. Additionally, A vaccine has some metadata that is strored in the blockchain which includes
- Manufacturer id
- Is_valid
- Is_injected
- Owner_history

The owner history enables the parties to track the whole vaccine owners . This provides full clarity and auditability to any party accessing the network.

*Figure 3: Sample vaccine metadata*

The manufacturer registers the vaccine in the chain by the *register_*vaccine function call and sets the metadata. Vaccines are assumed to be first admitted by the Authority and checked for the validity and conformity with the country policies. If It is confirmed by the authority, It will enter the supply chain by the *confirm_authority* function call that can only be called by the *Authority*. This is the point where the vaccine is given a valid flag.

After the confirmation by the *Authority*, Any change of vaccine possession is done via a process called *Handover*. *Handover* process is initiated by the party currently possessing the vaccine and with a *Handover_request* function call. This request will change the *ownership* state of the vaccine to a non-empty *next_owner* state, which means the vaccine is